\documentclass[twocolumn,showpacs,preprintnumbers,prb]{revtex4}
\usepackage{graphicx}
\usepackage{dcolumn}
\usepackage{bm}

\begin{document}
\preprint{}

\title{Effects of hole-doping on the magnetic ground state and excitations\\
in the edge-sharing CuO$_2$ chains of Ca$_{2+x}$Y$_{2-x}$Cu$_5$O$_{10}$}
\author{M.~Matsuda}
\author{K.~Kakurai}
\affiliation{Advanced Science Research Center,
Japan Atomic Energy Research Institute, Tokai, Ibaraki 319-1195, Japan}
\author{S.~Kurogi}
\author{K.~Kudo}
 \altaffiliation[Present address: ]{Institute for Materials Research, Tohoku University, Sendai 980-8577, Japan.}
\author{Y.~Koike}
\affiliation{
Department of Applied Physics, Tohoku University, Sendai 980-8579, Japan}
\author{H.~Yamaguchi}
\author{T.~Ito}
\author{K.~Oka}
\affiliation{
National Institute of Advanced Industrial Science and Technology,
Tsukuba, Ibaraki 305-8568, Japan}
\date{\today}

\begin{abstract}
Neutron scattering experiments were performed on the undoped and hole-doped Ca$_{2+x}$Y$_{2-x}$Cu$_5$O$_{10}$, which consists of ferromagnetic edge-sharing CuO$_2$ chains. It was previously reported that in the undoped Ca$_2$Y$_2$Cu$_5$O$_{10}$ there is an anomalous broadening of spin-wave excitations along the chain, which is caused mainly by the antiferromagnetic interchain interactions [Matsuda $et$ $al.$, Phys. Rev. B 63, 180403(R) (2001)]. A systematic study of temperature and hole concentration dependencies of the magnetic excitations shows that the magnetic excitations are softened and broadened with increasing temperature or doping holes irrespective of $Q$ direction. The broadening is larger at higher $Q$. A characteristic feature is that hole-doping is much more effective to broaden the excitations along the chain. It is also suggested that the intrachain interaction does not change so much with increasing temperature or doping although the anisotropic interaction and the interchain interaction are reduced. In the spin-glass phase ($x$=1.5) and nearly disordered phase ($x$=1.67) the magnetic excitations are much broadened in energy and $Q$. It is suggested that the spin-glass phase originates from the antiferromagnetic clusters, which are caused by the hole disproportionation.
\end{abstract}
\pacs{75.40.Gb, 75.30.Ds, 75.10.Jm}

\maketitle

\section{Introduction}
Doping dependence in the strongly correlated transition metal oxides has been studied extensively because various interesting phenomena such as high-$T\rm_c$ superconductivity or colossal magnetoresistance occur. An interesting phenomenon is that the doped holes can give rise to a charge ordering which also affects the magnetic ground state. The stripe order in La$_{2-x}$Sr$_x$NiO$_4$ and the underdoped La$_{2-x}$Sr$_x$CuO$_4$, in which charge density wave and spin density wave coexist, is an example.~\cite{tran1,tran2} In the edge-sharing CuO$_2$ chains in Sr$_{14}$Cu$_{24}$O$_{41}$, in which copper spins are coupled by the nearly 90$^\circ$ Cu-O-Cu interaction, the singlet dimers weakly coupled two-dimensionally are caused by the charge ordering of the doped holes.~\cite{ecc,regnault,matsu8}

Ca$_{2+x}$Y$_{2-x}$Cu$_5$O$_{10}$ system \cite{davis1,davis2,hayashi} is an candidate in which hole-doping effect on the magnetic ground state and excitations can be studied. A schematic structure of the edge-sharing CuO$_2$ chains is shown in Fig. 1. As reported by Hayashi, Batlogg, and Cava, this system does not show insulator-to-metal transition even when holes are doped by 40\%.~\cite{hayashi} Magnetically, the end-material Ca$_2$Y$_2$Cu$_5$O$_{10}$, which has no holes, shows an antiferromagnetic ordering of the Cu$^{2+}$ moment below 29.5 K with ferromagnetic coupling along the chain \cite{matsuda0,fong}. The ordered moment of Cu$^{2+}$ is ~0.9$\mu\rm_B$ at low temperatures, similar to the full magnetic moment of the free Cu$^{2+}$ ion.
Even though the magnetic interaction is ferromagnetic along the chain, in which quantum fluctuations are considered to be less prominent, a remarkable property in this system is that the magnetic excitation peak width in energy becomes broader with increasing $Q$ along the chain. From numerical calculations, it was revealed that magnetic excitations from ferromagnetic chains are considerably affected when a finite antiferromagnetic interchain coupling exists or frustration is introduced between the nearest-neighbor (NN) and next-nearest-neighbor (NNN) interactions in the chain.~\cite{mizuno2}
With hole-doping the long-range ordering is destroyed above $x$=1.3 and a spin-glass behavior appears. In the spin-glass region magnetic susceptibility measurements show a difference between field-cooling and zero-field-cooling processes.~\cite{kudo}
With more hole-doping main magnetic interaction becomes antiferromagnetic.~\cite{kudo,chabot,kurogi} It is also suggested that a spin gap originating from the singlet dimers appears.~\cite{kurogi} Therefore, it is interesting to study the hole-doping dependence of the magnetic ground state and also the anomalous excitations in this system. Especially, the charge ordering suggested for the Ca$_{2+x}$Y$_{2-x}$Cu$_5$O$_{10}$~\cite{chabot,kurogi} and the related system Ca$_{1-x}$CuO$_2$~\cite{dolinsek,hiroi,isobe} should be tested from a microscopic point of view.
\begin{figure}
\includegraphics[width=8cm]{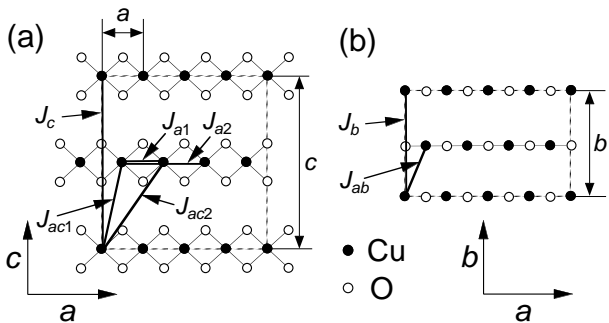}
\caption{Structure of the edge-sharing CuO$_2$ chains in the $ac$ plane (a) and in the $ab$ plane (b) in Ca$_{2+x}$Y$_{2-x}$Cu$_5$O$_{10}$. It is noted that oxygen ions are located at $z\sim\pm$0.125 in (b). The Cu$^{2+}$ moments align ferromagnetically along the chain ($a$ axis) with the propagation vector $k$=[001] below $T\rm_N$ in Ca$_{2+x}$Y$_{2-x}$Cu$_5$O$_{10}$ ($x\le$1.3). The Cu$^{2+}$ moments point along the $b$ axis. $J_{a1}$, $J_b$, $J_c$, $J_{ac1}$ and $J_{ab}$ are NN couplings along the $a$ (chain), $b$, $c$, ($\frac{1}{2}$,0,$\frac{1}{2}$), and ($\frac{1}{2}$,$\frac{1}{2}$,0) directions, respectively. $J_{a2}$ is a NNN coupling along the $a$ axis. $J_{ac2}$ is a coupling along ($\frac{3}{2}$,0,$\frac{1}{2}$).}
\label{fig1}
\end{figure}

In this study we performed neutron scattering experiments in single crystals of Ca$_{2+x}$Y$_{2-x}$Cu$_5$O$_{10}$ ($0\le x\le$1.67). The magnetic ground state was determined as shown in Fig. 2. The long-range magnetic order disappears around $x\sim$1.4. In the crystal with $x$=1.5 a broad transition to short-range ordered magnetic phase is found below $\sim$15 K with rather sharp development below 12 K. This probably originates from the spin-glass behavior expected from the magnetic susceptibility measurements.~\cite{kudo} In the crystal with $x$=1.67 the magnetic ground state is considered to be nearly disordered although a minor spin-glass phase was observed. A systematic study of temperature and hole concentration dependencies of the magnetic excitations shows that the magnetic excitations are softened and broadened with increasing temperature or doping holes. Hole-doping is much more effective to broaden the excitations along the chain. It was found that magnetic excitations are not resolution-limited even around the zone center in a doped sample ($x$=1.0) with a hole concentration of 20\%, which shows a long-range ordering.
In the spin-glass phase ($x$=1.5) and nearly disordered phase ($x$=1.67) the magnetic excitations are even more broadened. The magnetic excitations originating from the antiferromagnetic correlations in the chain or singlet dimers observed by magnetic susceptibility measurements were not observed although there is a possibility of a partial charge ordering, which gives rise to the spin-glass behavior.
\begin{figure}
\includegraphics[width=5cm]{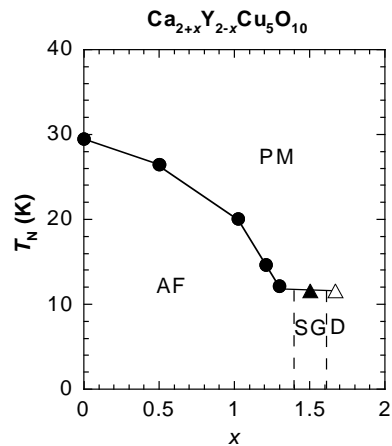}
\caption{Phase diagram of Ca$_{2+x}$Y$_{2-x}$Cu$_5$O$_{10}$. The transition temperatures are determined from the temperature dependence of the (003) magnetic peak intensity. The circles represent the transition temperatures of the long-range ordering. The triangles represent the transition temperatures of the quasi-static short-range ordering. The open triangle shows that the magnetic state is minor. PM, AF, SG, and D stand for paramagnetic, antiferromagnetic, spin-glass, and nearly disordered phases, respectively.}
\label{fig2}
\end{figure}

\section{Experimental Details}
The single crystals of Ca$_{2+x}$Y$_{2-x}$Cu$_5$O$_{10}$ ($x$=0, 0.5, 1.0, 1.2, 1.3, 1.5 and 1.67) were grown by the travelling solvent floating zone method. Typical dimensions of the rod shaped crystals were $\sim$6$\Phi\times$25 mm$^{3}$ for $x \le$1.3 and $\sim$4$\Phi\times$15 mm$^{3}$ for $x$=1.5 and 1.67. Detail of the crystal characterization is described elsewhere.~\cite{kurogi,kudo,yama}

The neutron-scattering experiments were carried out on the thermal neutron three-axis spectrometer TAS2 and the cold neutron three-axis spectrometer LTAS installed at the guide hall of JRR-3 at Japan Atomic Energy Research Institute. The fixed final neutron energy was 14.7 meV and 4 meV on TAS2 and LTAS, respectively. The typical horizontal collimator sequences for inelastic scattering measurements were guide-80'-S-40'-80' on TAS2 and guide-80'-S-80'-open on LTAS, respectively. For elastic scattering measurements the horizontal collimator sequences were guide-80'-S-80'-80' on TAS2 and guide-open-S-80'-open on LTAS, respectively. The single crystal, which was oriented in the $(H0L)$ or $(HK0)$ scattering plane, was mounted in a closed cycle refrigerator. Since the $Q$ resolution in the scattering plane is rather sharp, anomalous broadening of the magnetic excitations, which will be described in Sec. 3, is not expected from the resolution effect. The broad $Q$ resolution perpendicular to the scattering plane does not also broaden the magnetic excitations since the dispersion is almost flat.~\cite{matsuda3} Around the zone center the excitation peaks are slightly asymmetric and have a tail at higher energies due to the resolution effect, which is consistent with calculations.
\section{Magnetic excitations}
\subsection{Previous results in undoped Ca$_2$Y$_2$Cu$_5$O$_{10}$}
Matsuda $et$ $al$. previously performed inelastic neutron scattering experiments to study the spin-wave excitations in the antiferromagnetically ordered state in the end material Ca$_2$Y$_2$Cu$_5$O$_{10}$,~\cite{matsuda3} which contains no holes.
Applying the linear-spin-wave theory on a model Hamiltonian that includes uniaxial anisotropy, the dispersion of the magnetic excitations is given by
\begin{eqnarray}
\omega(\mbox{\boldmath $q$})
&=&\{[J_{a1}({\rm cos} q_a-1)+J_{a2}({\rm cos} 2q_a-1)\nonumber\\
&+&J_b({\rm cos} q_b-1)+J_c({\rm cos} q_c-1)\nonumber\\
&+&2J_{ab}({\rm cos}\frac{q_a}{2}{\rm cos}\frac{q_b}{2}-1)
+2J_{ac1}+2J_{ac2}-D]^2\nonumber\\
&-&(2J_{ac1}{\rm cos}\frac{q_a}{2}{\rm cos}\frac{q_c}{2}
+2J_{ac2}{\rm cos}\frac{3q_a}{2}{\rm cos}\frac{q_c}{2})^2\}^\frac{1}{2}.
\label{omega}
\end{eqnarray}
where $D=2J^{''}_{ac1}+2J^{''}_{ac2}-J^{''}_{a1}-J^{''}_{a2}-J^{''}_{b}-J^{''}_{c}-2J^{''}_{ab}$ and the magnetic interactions are shown in Fig. 1. $J^{''}$ represents the anisotropic exchange interaction $J^{z}-J^{x,y}$. These parameters were determined by fitting the dispersion relation as shown in Table 1. The nearest-neighbor interaction is ferromagnetic and fairy large. The interchain interaction along $c$ is antiferromagnetic, which is not negligible. It is noted that there is no frustration in the interactions. The anisotropic exchange interactions, which work to align the spins perpendicular to the chain plaquettes ($b$ axis), is not so small.~\cite{yu,tornow}
\begin{table*}
\caption{The magnetic coupling constants in Ca$_{2+x}$Y$_{2-x}$Cu$_5$O$_{10}$.}
\begin{ruledtabular}
\begin{tabular}{cccccccc}
$x$&$T$ (K)&$J_{a1}$ (meV)&$J_{ac1}$ (meV)&$D$ (meV)&$J_{ab}$ (meV)&$J_{b}$ (meV)&$J_{c}$ (meV)\\
\hline
0\footnote{Data from Ref. \onlinecite{matsuda3}.} & 7 & -6.9(1) & 1.494(3) & -0.262(3) & -0.030(3) & -0.061(6) & 0(fixed)\\
0 & 20 & -6.9(fixed) & 1.31(2) & -0.159(2) & 0(fixed) & - & 0(fixed)\\
0 & 25 & -6.9(fixed) & 1.16(2) & -0.104(2) & 0(fixed) & - & 0(fixed)\\
1.5 & 3 & -6.9(fixed) & 0.5(fixed) & -0.09(fixed) & 0(fixed) & - & 0(fixed)\\
1.67 & 3 & -6.9(fixed) & 0(fixed) & 0(fixed) & 0(fixed) & - & 0(fixed)\\
\end{tabular}
\end{ruledtabular}
\end{table*}

The most interesting feature is that the magnetic excitation peak width in energy becomes broader with increasing $Q$ along the chain although sharp excitations are observed around the zone center and perpendicular to the chain. Broadening of excitation peak width was reported also in a $S$=$\frac{1}{2}$ 1D Heisenberg ferromagnet CuCl$_2\cdot$DMSO.~\cite{satija} However, broadening is much more pronounced in Ca$_2$Y$_2$Cu$_5$O$_{10}$. From numerical calculations, it was revealed that the anomalous magnetic excitation spectra are caused mainly by the antiferromagnetic interchain interactions.

\subsection{Temperature dependence in Ca$_2$Y$_2$Cu$_5$O$_{10}$}
As the next step, we studied the temperature dependence of the spin-wave excitations in Ca$_2$Y$_2$Cu$_5$O$_{10}$. As shown in Fig. 3, excitation widths in energy are resolution-limited at the zone center along the chain and also at (0,0,1.3) perpendicular to the chain at 7 K. With increasing temperature, excitation widths become broader. As well as the broadening of the excitations, we also observed a softening of the excitations.
At higher temperatures we could only observe low energy excitations along the chain ($a$ axis) at $(H, 0, 1)$. Although the excitations at higher $Q$ and higher energies can be measured at $(H, 0, 3)$, the magnetic form factor decreases and the excitations are broadened so that distinct excitation peaks cannot be observed.
The solid lines are the results of fits to a convolution of the resolution function with a Lorentzian $\Gamma /(\Gamma^2+(\omega-\omega_0)^2)$, where $\Gamma$ and $\omega_0$ are inverse lifetime of magnetic excitations and excitation peak position, respectively.
The observed spectra are reproduced by the calculations quite well.
\begin{figure}
\includegraphics[width=8cm]{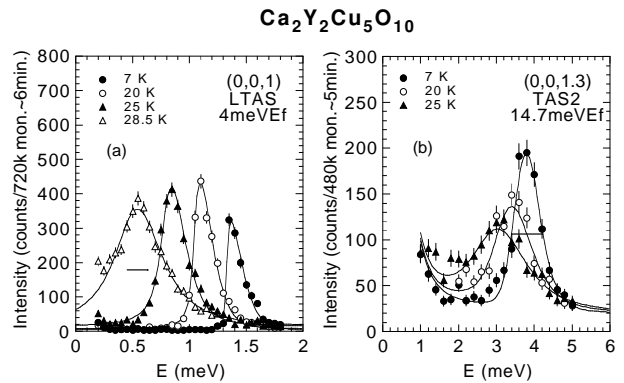}
\caption{Temperature dependence of magnetic excitations measured at (0, 0, 1) and (0, 0, 1.3) in Ca$_2$Y$_2$Cu$_5$O$_{10}$. The solid lines are the results of fits to a convolution of the resolution function with a Lorentzian. The thick horizontal bars represent the instrumental energy resolution. The background intensities in (b) at low energies below 2 meV are estimated from the data at 7 K.
}
\label{fig3}
\end{figure}

Figure 4 shows the $\omega$-$Q$ dispersion relations parallel and perpendicular to the chain as a function of temperature. The dispersion relations show almost a parallel shift to lower energies. We calculated the coupling constants by fitting the observed dispersion data to the dispersion relation Eq. (1). The curves are dispersion relations calculated with parameters as shown in Table 1. The observed dispersion relations are reproduced by the calculations reasonably well. It is noted that the dispersion along $l$ does not depend on $J_{a1}$ and $J_{ab}$ but $J_{ac1}$, $J_{ac2}$(=2$J_{ac1}$), $J_{c}$, and $D$. Therefore, $J_{ac1}$ and $D$ were first fitted with the dispersion along $l$ and then $J_{a1}$ was fitted with the dispersion along $h$. In the calculations $J_{a1}$ at 20 and 25 K was fixed at the fitted value at 7 K since number of the data points is limited. $J_{ab}$ at 20 and 25 K was fixed at zero since
the parameter is expected to be very small and also has a small effect on the dispersion along $h$. $J_{c}$ at 20 and 25 K was also fixed at zero as at 7 K.~\cite{matsuda3} The dispersion relations along $h$ and $l$ are independent of $J_{b}$.
These results suggest that with increasing temperature the intrachain interaction is not affected so much but the interchain interaction $J_{ac1}$ and the anisotropic exchange interaction $D$ decrease considerably.
\begin{figure}
\includegraphics[width=8cm]{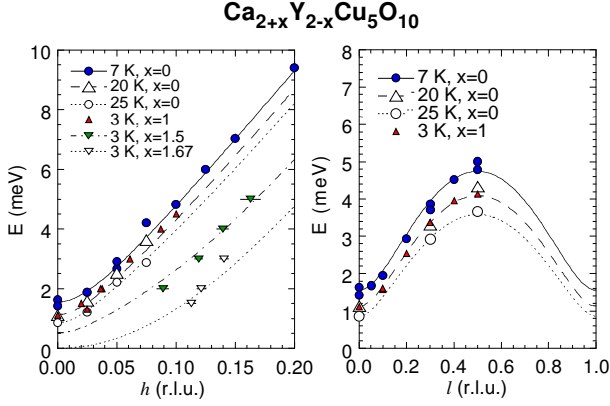}
\caption{(Color online) Temperature dependence of $\omega$-$Q$ dispersion relations for the edge-sharing CuO$_2$ chain in Ca$_2$Y$_2$Cu$_5$O$_{10}$ along the $a$ (chain) and $c$ axes. $\omega$-$Q$ dispersion relations in Ca$_{2+x}$Y$_{2-x}$Cu$_5$O$_{10}$ ($x$=1, 1.5, and 1.67) measured at 3 K are also shown. The curves represent the theoretical ones with the magnetic interactions shown in Table 1.
}
\label{fig4}
\end{figure}

Figure 5 shows the intrinsic peak width in energy ($\Gamma$) parallel and perpendicular to the chain as a function of temperature. The widths are resolution-limited or close to resolution-limited below $h\sim$0.1 along the chain and also up to the zone boundary perpendicular to the chain at 7 K. The widths increase with increasing temperatures throughout the whole Brillouin zone including around the zone center. The thermal effect is larger at larger $Q$ and higher energies irrespective of $Q$ direction.
\begin{figure}
\includegraphics[width=8cm]{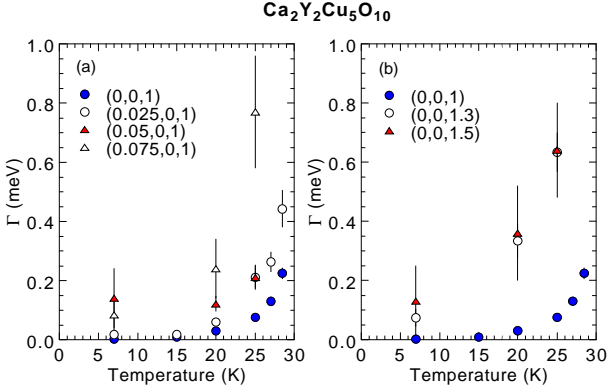}
\caption{(Color online) Temperature dependence of excitation width in energy ($\Gamma$) at various $Q$ positions along the $a$ (chain) and $c$ axes in Ca$_2$Y$_2$Cu$_5$O$_{10}$.
}
\label{fig5}
\end{figure}

\subsection{Hole-doping effect in doped Ca$_{2+x}$Y$_{2-x}$Cu$_5$O$_{10}$ ($x\le$1.3)}
Ca$_3$Y$_1$Cu$_5$O$_{10}$ has a small amount of holes ($\sim$20\%), which are localized. This sample shows a long-range antiferromagnetic ordering below $\sim$20.5 K. The magnetic structure is the same as in the undoped Ca$_2$Y$_2$Cu$_5$O$_{10}$. Figure 6 shows the typical neutron inelastic spectra of constant-$Q$ scans at $(H, 0, L)$ measured at 3 K on LTAS and TAS2.
The excitation peaks are broader than those in undoped Ca$_2$Y$_2$Cu$_5$O$_{10}$.
As shown in Fig. 6(a), the excitation peak in energy is broader than instrumental resolution even at the zone center and becomes broader at larger $Q$.
Figure 7 shows the typical neutron inelastic spectra of constant-$\omega$ scans at $(H, 0, 1)$ measured at 3 K on TAS2. A broad magnetic peak originating from the spin wave excitations was observed.
\begin{figure}
\includegraphics[width=8cm]{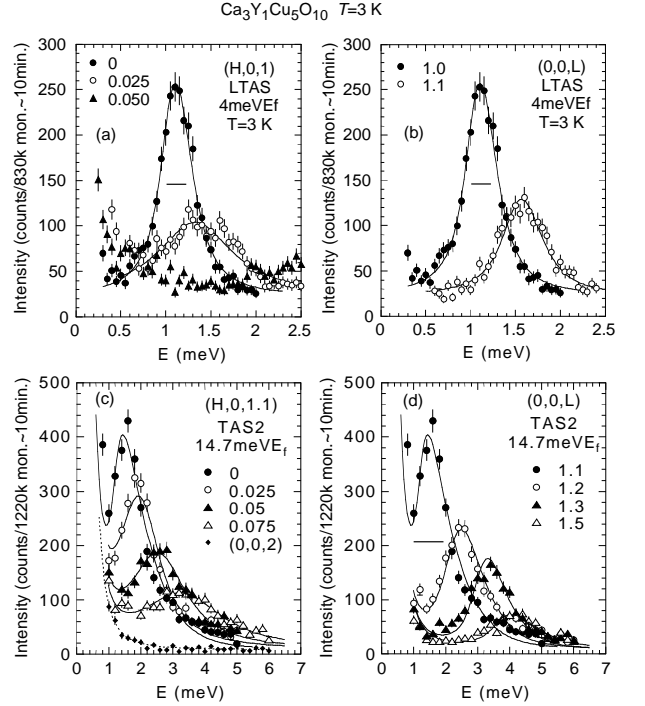}
\caption{Constant-$Q$ scans at $(H, 0, L)$ measured at $T$=3 K in Ca$_3$Y$_1$Cu$_5$O$_{10}$. The solid lines are the results of fits to a convolution of the resolution function with a Lorentzian. Note that background intensity is included in the fitting in (c) and (d). The dotted line represents the background intensity estimated from a constant-$Q$ scan at the zone boundary (0, 0, 2), where magnetic scattering is considered to be negligible. The thick horizontal bars represent the instrumental energy resolution.
}
\label{fig6}
\end{figure}
\begin{figure}
\includegraphics[width=8cm]{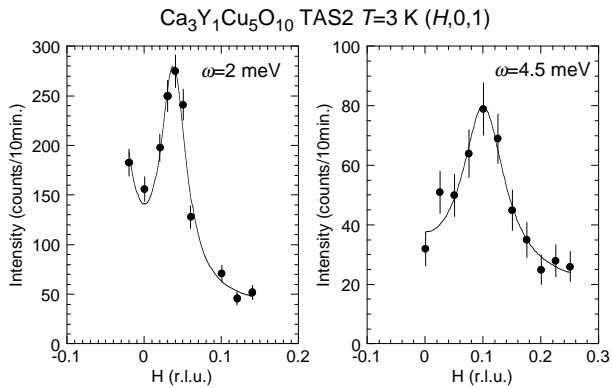}
\caption{Constant-$\omega$ scans at $(H, 0, 1)$ measured at $\omega$=2 and 4.5 meV and $T$=3 K in Ca$_3$Y$_1$Cu$_5$O$_{10}$. The solid lines are the results of fits to two Gausians centered at ($\pm h_0$, 0, 1). 
}
\label{fig7}
\end{figure}

The observed excitation energies along $h$ and $l$ are shown in Fig. 4. Along $l$ all the data are obtained from constant-$Q$ scans as shown in Figs. 6(b) and 6(d). On the other hand, along $h$ data are obtained from both constant-$Q$ and constant-$\omega$ scans as shown in Figs. 6(a) and 7. The spin-wave excitations are softened both along $h$ and $l$. The dispersion relations in Ca$_3$Y$_1$Cu$_5$O$_{10}$ at 3 K are similar to those observed at 20 K in undoped Ca$_2$Y$_2$Cu$_5$O$_{10}$, suggesting that the hole-doping and temperature affect the spin-wave excitations similarly.

Figure 8 shows $\Gamma$ along $h$ and $l$ in Ca$_{2+x}$Y$_{2-x}$Cu$_5$O$_{10}$ ($x$=0 and 1) at various temperatures. Because of the large incoherent scattering at (0, 0, 1) in Ca$_3$Y$_1$Cu$_5$O$_{10}$, the measurements at higher $Q$ along $h$ were performed at ($h$, 0, 1.1) on TAS2 as shown in Fig. 6(c). The magnetic excitation peaks around the zone center were measured on LTAS as shown in Fig. 6(a). As observed in undoped Ca$_2$Y$_2$Cu$_5$O$_{10}$ at low temperatures, $\Gamma$ broadens at higher $Q$ and higher energies in Ca$_3$Y$_1$Cu$_5$O$_{10}$. The $l$ dependence of $\Gamma$ in Ca$_3$Y$_1$Cu$_5$O$_{10}$ at 3 K is similar to that in Ca$_2$Y$_2$Cu$_5$O$_{10}$ at 20 K. This behaviour is consistent with the softening of the spin-wave excitations mentioned above. The most characteristic feature is that $\Gamma$ along the chain is much larger than that in Ca$_2$Y$_2$Cu$_5$O$_{10}$ at 20 K.
\begin{figure}
\includegraphics[width=8cm]{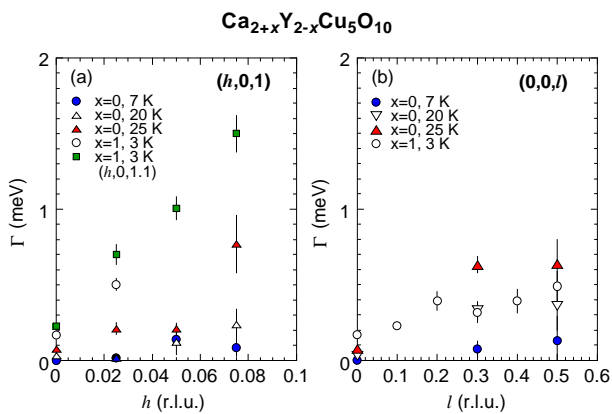}
\caption{(Color online) Excitation width in energy ($\Gamma$) along the $a$ (chain) and $c$ axes at various temperatures in Ca$_2$Y$_2$Cu$_5$O$_{10}$ and at 3 K in Ca$_3$Y$_1$Cu$_5$O$_{10}$.
}
\label{fig8}
\end{figure}

We also measured magnetic excitations in Ca$_{2+x}$Y$_{2-x}$Cu$_5$O$_{10}$ ($x$=1.2 and 1.3). The softening and broadening in the magnetic excitations are much larger than in Ca$_3$Y$_1$Cu$_5$O$_{10}$.

\subsection{Spin-glass Ca$_{3.5}$Y$_{0.5}$Cu$_5$O$_{10}$ and nearly disordered Ca$_{3.67}$Y$_{0.33}$Cu$_5$O$_{10}$}
In Ca$_{2+x}$Y$_{2-x}$Cu$_5$O$_{10}$ ($x$=1.5 and 1.67) magnetic susceptibility measurements show a difference between field-cooling and zero-field-cooling processes,~\cite{kudo} suggesting a spin-glass behavior at low temperatures. Neutron scattering experiments were carried out in these compounds. Broad magnetic peaks were observed at the same positions, where sharp magnetic Bragg peaks are observed in Ca$_{2+x}$Y$_{2-x}$Cu$_5$O$_{10}$ ($x\le$1.3).~\cite{matsuda0,fong} Figure 9(a) shows the temperature dependence of the magnetic elastic intensity measured at (0, 0, 3) in Ca$_{3.5}$Y$_{0.5}$Cu$_5$O$_{10}$. The intensity depends on the energy resolution of the incident neutron beam. The transition temperature becomes lower when it is measured with lower energy neutrons, in which energy window is narrower (0.2 and 0.9 meV in incident neutron energies of 4 and 14.7 meV, respectively). The transition temperature $\sim$12 K is about factor of 2 higher than that determined from the susceptibility measurements.~\cite{kudo} This is a typical property in the spin-glass phase, in which magnetic moments are fluctuating and magnetic correlations are short-ranged. In Ca$_{3.67}$Y$_{0.33}$Cu$_5$O$_{10}$ both the temperature and incident neutron energy dependencies are similar to those in Ca$_{3.5}$Y$_{0.5}$Cu$_5$O$_{10}$ with much reduced intensity, suggesting that the spin-glass phase is minor and the most of the moments are disordered even at low temperatures.
\begin{figure}
\includegraphics[width=8cm]{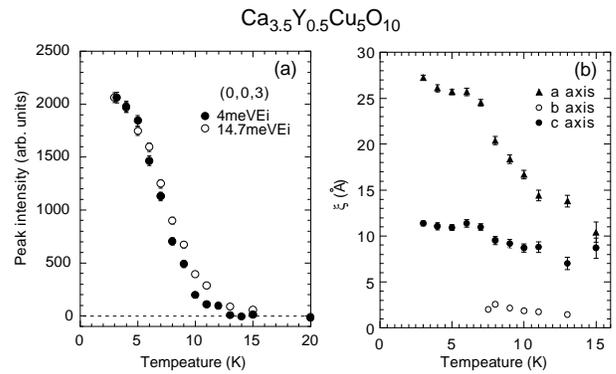}
\caption{Temperature dependence of the (003) magnetic peak intensity (a) and correlation lengths along the $a$, $b$ and $c$ axes (b) in Ca$_{3.5}$Y$_{0.5}$Cu$_5$O$_{10}$. Open and filled circles represent the data measured with incident neutron energies of 14.7 meV ($\Delta\omega\sim$0.9 meV) and 4 meV ($\Delta\omega\sim$0.2 meV), respectively. Background intensity measured at high temperatures is subtracted. The intensities measured with two different conditions are normalized at 3K.
}
\label{fig9}
\end{figure}

Figures 10 shows neutron elastic scattering spectra observed around (0, 0, 3) with an incident neutrons of 14.7 meV in Ca$_{3.5}$Y$_{0.5}$Cu$_5$O$_{10}$. The peak widths are broad along all the $Q$ directions. The solid lines are the results of fits to a convolution of the resolution function with a three-dimensional Lorentzian squared $1/[1+\xi_a^2(q_a-q_{0a})^2+\xi_b^2(q_b-q_{0b})^2+\xi_c^2(q_c-q_{0c})^2]^2$, where $\xi$ and $q_0$ are magnetic correlation length and peak position, respectively. The observed spectra are reproduced by the calculations reasonably well.
\begin{figure}
\includegraphics[width=6cm]{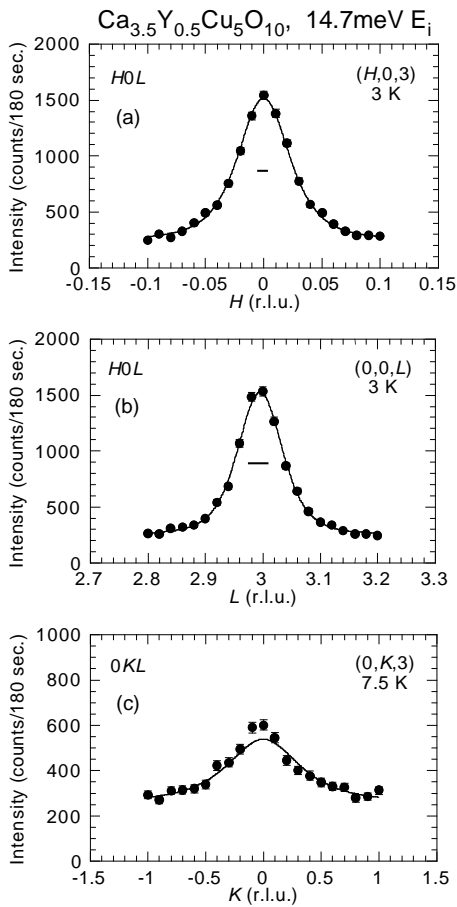}
\caption{Magnetic elastic neutron scattering spectra at (0, 0, 3) in Ca$_{3.5}$Y$_{0.5}$Cu$_5$O$_{10}$. The solid lines are the results of fits to a convolution of the resolution function with a Lorentzian squared. The thick horizontal bars represent the instrumental $Q$ resolution. The resolution is not shown in (c) along the $k$ direction since it is negligibly small.
}
\label{fig10}
\end{figure}

The correlation lengths along the $a$, $b$, and $c$ axes determined from the fittings are shown in Fig. 9(b). They increase gradually with decreasing temperature. The magnetic correlations are highly anisotropic and the longest ($\sim$28\AA) along the chain and the shortest ($\sim$2\AA) perpendicular to the plane of the CuO$_2$ unit. This result is consistent with the anisotropic magnetic interactions. In Ca$_{3.5}$Y$_{0.5}$Cu$_5$O$_{10}$ with a hole concentration of 30\%, an averaged spin cluster size is $\sim$6\AA\ along the chain. The correlation length is much larger than the cluster size.

Figure 11 shows the typical neutron inelastic spectra of constant-$Q$ scans at $(H0L)$ in Ca$_{2+x}$Y$_{2-x}$Cu$_5$O$_{10}$ ($x$=1.5 and 1.67) measured at 3 K on LTAS and TAS2. The excitation peaks are much broader than those in Ca$_3$Y$_1$Cu$_5$O$_{10}$ and it is very difficult to observe a distinct excitation peak in energy both along $h$ and $l$. However, as shown in Fig. 12, broad magnetic peaks originating from the spin wave excitations along $h$ were observed in constant-$\omega$ scans although no distinct peak was observed even in constant-$\omega$ scans along $l$. The peak positions are plotted in Fig. 4(a). In the calculations of the dispersion relation along $h$ in the $x$=1.5 and 1.67 samples all the parameters are fixed at the values shown in Table 1 since the number of the data points is limited. $J_{a1}$ in the both samples is fixed at the value of that in Ca$_2$Y$_2$Cu$_5$O$_{10}$ at 7 K. In Ca$_{3.5}$Y$_{0.5}$Cu$_5$O$_{10}$ $J_{ac1}$ and $D$ are fixed at the values one third of those in Ca$_2$Y$_2$Cu$_5$O$_{10}$ at 7 K. These values are fixed at zero in Ca$_{3.67}$Y$_{0.33}$Cu$_5$O$_{10}$. The assumption of $J_{ac1}$ for both samples is reasonable since no distinct peak was observed along $l$ in both constant-$Q$ and constant-$\omega$ scans. The observed dispersion relations are reproduced by the calculations reasonably well.
The above analysis suggests that the intrachain interaction is not affected so much but the anisotropic exchange interaction decreases considerably with hole doping. This is consistent with the result of the magnetization measurements that the saturated Cu$^{2+}$ moment does not change but spin-flop transition field decreases with hole doping.~\cite{kudo}

In these samples we also measured magnetic excitations below 8 meV around antiferromagnetic positions along the chain ($\frac{1}{2}$, 0, 1) and ($\frac{1}{2}$, 0, 2) and also around antiferromagnetic positions with two times periodicity ($\frac{1}{4}$, 0, 1), ($\frac{3}{4}$, 0, 1), ($\frac{1}{4}$, 0, 2), and ($\frac{3}{4}$, 0, 2) as in the Sr$_{14}$Cu$_{24}$O$_{41}$ system.~\cite{ecc,regnault,matsu8} The elastic peak at ($\frac{1}{2}$, 0, 1), where magnetic signal was observed in Ca$_{0.83}$CuO$_2$,~\cite{meijer,note2} was also measured. However, no distinct magnetic signal was observed. This suggests that the antiferromagnetic correlations or singlet dimers observed by other measurements~\cite{dolinsek,meijer,hiroi,kurogi} just exist as a minor phase in these samples even if they exist.
\begin{figure}
\includegraphics[width=8cm]{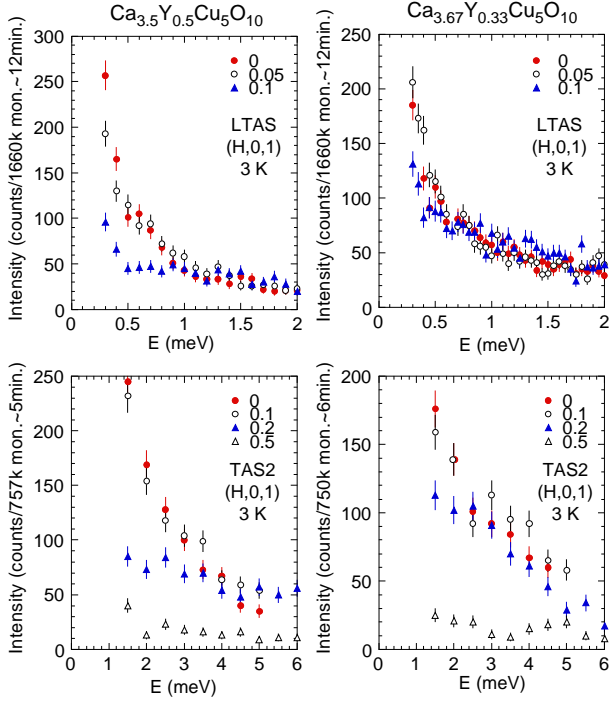}
\caption{(Color online) Constant-$Q$ scans at $(H, 0, L)$ measured at $T$=3 K in Ca$_{2+x}$Y$_{2-x}$Cu$_5$O$_{10}$ ($x$=1.5 and 1.67).}
\label{fig11}
\end{figure}
\begin{figure}
\includegraphics[width=8cm]{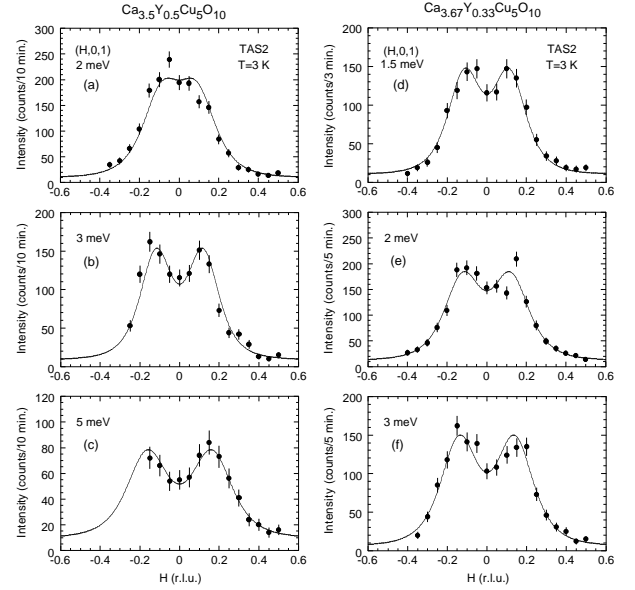}
\caption{Constant-$\omega$ scans at $(H, 0, 1)$ measured at various energies and $T$=3 K in Ca$_{2+x}$Y$_{2-x}$Cu$_5$O$_{10}$ ($x$=1.5 and 1.67). The solid lines are the results of fits to two Gausians centered at ($\pm h_0$, 0, 1).
}
\label{fig12}
\end{figure}
 
\section{Discussion}
In Ca$_{2+x}$Y$_{2-x}$Cu$_5$O$_{10}$ magnetic excitations are softened and broadened with increasing temperature, which is similar to the hole-doping effect qualitatively. However, the hole-doping is much more effective to broaden the excitations along the chain. This behavior is probably explained as follows. The spin of the doped-hole is coupled with the Cu$^{2+}$ moment to form the Zhang-Rice (ZR) singlet~\cite{ZR}. The ZR singlet cuts the magnetic bond mostly along the chain so that it disturbs magnetic excitations along the chain. Since doped holes affect the excitation width along the chain very effectively, the hole-doping effect is larger than that of the thermal effect. It should be pointed out that the long-range magnetic ordering persists even when the spin-wave excitations are considerably broadened.

With hole-doping $T\rm_N$ decreases slowly. The hole concentration at which long-range magnetic ordering disappears is $\sim$28\%. This is even comparable with the percolation threshold for $S=\frac{1}{2}$ square-lattice Heisenberg antiferromagnet, 40.725\%.~\cite{aharony,newman,vajk} Since $J_{a1}$ is much larger than $J_{ac1}$, it may seem that Ca$_{2+x}$Y$_{2-x}$Cu$_5$O$_{10}$ is a quasi-one-dimensional system. However, $J_{ac2}$ also works as antiferromagnetic interchain coupling along the $c$ axis. Furthermore, the number of interaction bonds of $J_{ac1}$ and $J_{ac2}$ is twice as much as that of $J_{a1}$. Therefore, the end material Ca$_2$Y$_2$Cu$_5$O$_{10}$ is a system between being one- and two-dimensional. The long-range magnetic order probably disappears when with hole-doping the interchain couplings become small compared to the intrachain coupling. Experimentally, magnetic order disappears completely in Ca$_{3.67}$Y$_{0.33}$Cu$_5$O$_{10}$, in which interchain couplings are negligibly small. 

The interchain and the anisotropic exchange interactions are reduced considerably with increasing temperature or doping holes although the intrachain interaction does not change so much. Therefore, it is considered that above $T\rm_N$ the magnetic interactions are well described by the one-dimensional Heisenberg model.
We first expected that the one-dimensional spin-wave excitations can be observed even above $T\rm_N$ since the NN interaction along the chain is -6.9 meV (=80 K) in Ca$_2$Y$_2$Cu$_5$O$_{10}$. However, with increasing temperature magnetic excitations are damped around $T\rm_N$ so that distinct excitation peaks cannot be observed above $T\rm_N$.

As described in Sec. 1, it was reported that with hole-doping the long-range ordering is destroyed above $x$=1.3 and a spin-glass behavior appears in the $x$=1.5 sample.~\cite{kudo} Here, we consider how the spin-glass phase is caused. As described above, the magnetic interactions do not show frustration in the $x$=1 sample and probably in the $x$=1.5 sample. In addition, the magnetic correlations in the spin-glass phase is commensurate and the same as that in the undoped Ca$_2$Y$_2$Cu$_5$O$_{10}$. Therefore, it is difficult to understand how the spin-glass phase appears in the absence of frustrating interactions. It may be possible that doped holes are not localized randomly but partially ordered \cite{note} and as the results short-ranged magnetic clusters, which give rise to a cluster spin-glass behavior, are formed although the ordering is not long-ranged. This is consistent with the fact that the correlation length along the chain is larger than the average spin cluster size as described in Sec. 3D. It is interesting that the spin-glass behavior appears in a very narrow region $x\sim$1.5. This is probably because finite interchain couplings are needed to form finite-sized spin clusters which can realize the cluster-spin-glass behavior.
It is theoretically predicted that in two-dimensional Ising system with mobile defects finite sized clusters are formed.~\cite{selke,holtschneider}
With more hole-doping the magnetic correlations along the chain change from being ferromagnetic to antiferromagnetic~\cite{hayashi,meijer,chabot,kurogi} and a dimerized state finally appears.~\cite{dolinsek,hiroi,kurogi} There might be a drastic change in magnetic interactions above $x$=1.7 probably driven by a charge ordering as in the Sr$_{14}$Cu$_{24}$O$_{41}$ system.~\cite{ecc,regnault,matsu8}

It is interesting to compare the results in Ca$_{2+x}$Y$_{2-x}$Cu$_5$O$_{10}$ with those in related system La$_5$Ca$_9$Cu$_{24}$O$_{41}$ that also consists of edge-sharing CuO$_2$ chains. La$_5$Ca$_9$Cu$_{24}$O$_{41}$ shows antiferromagnetic ordering below 10.5 K.~\cite{matsuda2} In this compound, the characteristic features are that the magnetic interactions are frustrating and that broadening of the excitation peak is much more enhanced than in Ca$_{2+x}$Y$_{2-x}$Cu$_5$O$_{10}$.~\cite{matsuda4} The latter originates from the combined effects of frustrating interactions \cite{mizuno2} and disorder introduced by a small amount of holes ($\sim$10\%) and by a slight structural distortion in La$_5$Ca$_9$Cu$_{24}$O$_{41}$.

In summary, a systematic study of temperature and hole concentration dependencies of the magnetic excitations in Ca$_{2+x}$Y$_{2-x}$Cu$_5$O$_{10}$ shows that the magnetic excitations are softened and broadened with increasing temperature or doping holes. It was also suggested that the intrachain interaction does not change so much with increasing temperature or doping although the anisotropic interaction and the interchain interaction are reduced. In the spin-glass phase ($x$=1.5) and nearly disordered phase ($x$=1.67) the magnetic excitations are much broadened in energy and $Q$. Neither antiferromagnetic correlations nor singlet dimers caused by a charge ordering coexist with the spin-glass or the nearly disordered phase although there is a possibility of a partial charge ordering, which gives rise to the spin-glass behavior.



\begin{thebibliography}{}
\bibitem{note2}It is noted that the crystal axes are different in two compounds. The $a$ and $b$ axes in Ca$_2$Y$_2$Cu$_5$O$_{10}$ correspond to the $b$ and $a$ axes in Ca$_{0.83}$CuO$_2$, respectively.
\bibitem{note}The configuration of the magnetic clusters may be similar to that in Fig. 3 in Ref. \onlinecite{chabot}. However, the averaged magnetic cluster size is 10$a\times c$
in Ca$_{3.5}$Y$_{0.5}$Cu$_5$O$_{10}$. Furthermore, there is no long-range ordering of the clusters.
\end{thebibliography}
\end{document}